\documentclass[prd, twocolumn, nofootinbib]{revtex4}

\usepackage{epsfig} 

\newcommand{\beq}{\begin{equation}}
\newcommand{\eeq}{\end{equation}}
\newcommand{\beqa}{\begin{eqnarray}}
\newcommand{\eeqa}{\end{eqnarray}}
\newcommand{\om}{\Omega_m}

\def\ga{\mathrel{\mathpalette\fun >}}
\def\fun#1#2{\lower3.6pt\vbox{\baselineskip0pt\lineskip.9pt
  \ialign{$\mathsurround=0pt#1\hfil##\hfil$\crcr#2\crcr\sim\crcr}}}

\begin{document} 

\title{Cosmic Growth History and Expansion History} 
\author {Eric V.\ Linder} 
\affiliation{Physics Division, Lawrence Berkeley Laboratory, 
Berkeley, CA 94720} 

\date{\today} 

\begin{abstract} 
The cosmic expansion history tests the dynamics of the global evolution 
of the universe and its energy density contents, while the cosmic growth 
history tests the evolution of the inhomogeneous part of the energy 
density.  Precision comparison of the two histories can distinguish the 
nature of the physics responsible for the accelerating cosmic expansion: 
an additional smooth component -- dark energy -- or a modification of 
the gravitational field equations.  With the aid of a new fitting formula 
for linear perturbation growth accurate to 0.05-0.2\%, we separate out 
the growth 
dependence on the expansion history and introduce a new growth index 
parameter $\gamma$ that quantifies the gravitational modification. 
\end{abstract} 

%\keywords{dark energy --- cosmology:observations --- 
%cosmology: theory}

\maketitle 

\section{Introduction} \label{sec.intro}

Acceleration of the cosmic expansion reveals fundamentally new physics 
missing from our picture of the universe, yet key for the understanding 
of the recent and present history and the fate of the universe.  Furthermore, 
this new physics tells us that our standard models of gravitation and 
particle physics may be woefully incomplete.  The acceleration may lead 
us to insights about new high energy physics and the nature of the quantum 
vacuum, or about gravitation beyond Einstein's general relativity.  Perhaps 
most exciting would be a signal that both are involved, providing clues 
to a theory of quantum gravity. 

The first scenario includes physical components such as the cosmological 
constant, dynamical scalar field models, or effective potentials from 
string theory.  The second scenario includes extensions of the 
Einstein-Hilbert action, e.g.\ to higher derivative theories, scalar-tensor 
theories, generalized functions of the Ricci scalar, theories of 
supergravity or quantum gravity, and infrared modifications of gravity 
such as from hidden spacetime dimensions.  We can say that searching for 
the nature of the accelerating expansion is seeking to answer one or 
the other question: ``Does nothing weigh something?'' or ``Is nowhere 
somewhere?'' 

To distinguish the many different theoretical possibilities requires 
accurate observations of the cosmic expansion history, but even this 
will leave some degeneracies between explanations.  Models with different 
physical origins but the same global expansion properties could not be 
separated.  Fortunately, the overall smooth characteristics of the universe 
are not the only observables.  The energy density contents have evolved 
from the hot, dense, smooth state of the early universe to a relatively 
cool, diffuse, and in the case of matter, clustered state.  While the 
first two properties are purely due to the expansion of the zeroth order, 
homogeneous universe, being qualitatively kinematical, the last property 
arises from the perturbed, inhomogeneous universe, being intrinsically 
dynamical\cite{tegmark}.  
The growth of large scale structure in the universe 
provides an important companion test, and the cosmic expansion history 
and growth history together provide discernment of the nature of the new 
accelerating physics. 

In \S\ref{sec.exp} we discuss the expansion history and the effective 
equation of state entering the acceleration.  The growth 
of linear perturbations in the matter component in a generalized cosmological 
model is reviewed in \S\ref{sec.groint}.  The growth equation is 
extended in \S\ref{sec.groext} to allow other theories of gravitation 
besides general relativity, and formal solutions given.  For practical 
use in constraining models by observational data we introduce a highly 
accurate fitting formula in \S\ref{sec.fit} and apply it to a braneworld 
gravity model and models with coupling between the matter and dark energy 
density.  We present the conclusions in 
\S\ref{sec.concl}. 

\section{Expansion History} \label{sec.exp} 

The expansion history of the universe is a key quantity in cosmology, 
appearing directly in the metric in the form $a(t)$.  Kinematically, 
this is all that is needed to define distances and volumes (together 
with the spatial curvature constant $k$, which we take to be zero, 
though this does not affect the form of the following discussion). 
To evaluate the distances for a specific cosmology, dynamics or 
equations of motion from the gravity theory are required, together 
with information on the energy density contents.  The expansion history, 
together with the amount of clustering matter and any interactions 
of the matter with other components, is the central ingredient as well 
in the growth of matter perturbations. 

The Friedmann expansion equation in terms of the Hubble parameter 
$H=\dot a/a$ is 
\beq 
H^2=(8\pi G/3)\sum_i \rho_i, 
\eeq 
where we sum over all components of the energy density.  Since we are 
especially interested in the matter component, e.g.\ since we are 
positive it exists and since we will later examine its growth into 
large scale structure, it is convenient to separate it out from the sum. 
Then in terms of dimensionless energy density we can write 
\beqa 
H^2/H_0^2&=&\om a^{-3}+\sum_{i'} \Omega_{i'}e^{3\int_a^1 
[1+w_{i'}(a')]da'/a'} \\ 
&=&\om a^{-3}+\delta H^2/H_0^2, 
\eeqa 
where the set $i'$ does not include matter, $\om+\sum_{i'}\Omega_{i'}=1$, 
and $w(a)$ is the equation of state of each component.  

Without imposing any physical interpretation on $\delta H^2$ as actually 
being due to an energy density component as opposed to a modification 
of the Friedmann expansion equation, we can define an effective 
``acceleration physics'' or ``dark energy'' equation 
of state \cite{linjen} (cf.\ \cite{staro}) 
\beqa 
w(a)&=&-1-\frac{1}{3}\frac{d\ln \delta H^2}{d\ln a} \\ 
&=&-\frac{1}{3}\frac{d\ln [\om(a)^{-1}-1]}{d\ln a},  
\eeqa 
writing $\om(a)= \om a^{-3}/(H/H_0)^2$. 

Volumes and distances are built up out of the conformal distance 
\beq 
\eta(a)=\int_a^1 (da'/a')(a'H)^{-1}=\int_0^z dz'/H. 
\eeq 
Models with the same expansion history will have the same distances 
and volumes.  Note that formally we could obtain the same expansion history 
for two models by keeping their $\om$ the same and matching their $w(a)$, 
or by allowing different $\om$ and compensating for this in $w(a)$. 
Since the latter case corresponds to misestimating the matter density 
rather than any new physics, we do not consider it further.  

While the definition of an effective equation of state in terms of the 
expansion history is powerful, allowing different models to be talked 
about with a common language and treated in a model independent parameter 
space, this feature is also a bug.  Measurements of the expansion history, 
through distances and volumes to arbitrary precision, will not be able to 
distinguish different physical origins for the same expansion behavior. 
This is where the growth history comes in. 

\section{Growth of Matter Density Perturbations} \label{sec.groint} 

The universe has not remained homogeneous on all scales 
since its early, essentially smooth state.  While the largest volumes 
can still be treated as homogeneous and isotropic Robertson-Walker 
universes, smaller scale evolution must take into account perturbations 
to the metric in the form of gravitational potentials. 

Note that recent speculation \cite{rocky1,rocky2} about the interaction 
of these potentials to affect significantly even the global expansion 
seems misplaced; investigation of a realistic inhomogeneous universe 
metric by Jacobs, Linder, \& Wagoner \cite{jlw1,jlw2} derived a Green 
function solution for the potential.  This ``post-Newtonian'' solution 
corrects the Newton-Poisson equation and shows that no infrared 
divergences exist in the potential, rather a suppression as the Hubble 
scale is approached.  The Appendix summarizes the effects. 

Using the perturbed equations of motion for the gravity theory, 
one can derive the growth of density perturbations. 
Concentrating on perturbations in the matter density 
$\delta=\delta\rho_m/\rho_m$, assuming all other components are smooth, 
within general relativity the time evolution is 
\beq 
\ddot\delta+2H\dot\delta-4\pi G\rho_m\delta=0. \label{eq.grot}
\eeq 
The physical interpretation is very simple: the perturbations grow 
according to a source term involving the amount of matter able to cluster 
and are restricted by a friction term, or Hubble drag, arising from 
the expansion of the universe.  General discussion of the physics 
dependence on the expansion rate is in \cite{linjen}. 

It is convenient to study the growth evolution in terms of the 
expansion scale $a$ or characteristic (e-fold) scale $\ln a$, rather 
than time $t$.  Since the pure matter universe solution has $\delta\sim a$, 
it is also useful in studying the dark energy to divide out this behavior 
and switch to the growth variable $g=\delta/a$.  
Finally, since we will be interested in modifications of gravity, we 
hereafter normalize $G$ by Newton's constant, i.e.\ where $G$ appears in 
equations it stands for $G/G_{\rm Newton}$. 

Denoting derivatives with 
respect to $\ln a $ as primes, we have: 
\beqa 
g''&+&\left[4+\frac{1}{2}(\ln H^2)'\right]g' \nonumber\\ 
&+&\left[3+\frac{1}{2}(\ln H^2)'-\frac{3}{2}G\om(a) 
\right]g=0, \label{eq.groh} \\ 
g''&+&[3-q]g' \nonumber\\ 
&+&\left[2-q-\frac{3}{2}G\om(a)\right]g=0, \label{eq.groq} \\
g''&+&\left[\frac{5}{2}-\frac{3}{2}w(a)\Omega_w(a)\right]g' \nonumber\\ 
&+&\frac{3}{2}[1-w(a)]G\Omega_w(a)\,g=0, \label{eq.grow} \\
g''&+&\left[\frac{5}{2}-\frac{1}{2}(\ln\om(a))'\right]g' \nonumber\\ 
&+&\left[\frac{3}{2}-\frac{1}{2}(\ln\om(a))'-\frac{3}{2}G\om(a)
\right]g=0. \label{eq.groom} 
\eeqa 

All these forms are equivalent.  They clearly show that the Hubble drag 
is increased, and hence growth is suppressed, for an accelerating universe, 
as the deceleration parameter $q=-a\ddot a/\dot a^2$ or $w$ become more 
negative.  We emphasize that they also 
demonstrate that within 
general relativity the linear theory growth factor depends purely on the 
expansion history, e.g.\ $H(a)$ or $w(a)$ or $\om(a)$ or 
$\Omega_w(a)=1-\om(a)$. 
So a discrepancy between the growth observed and that predicted based on 
an observed expansion history tests the theoretical framework and can 
point up modifications to the theory of gravity.

\section{Generalization to Gravitational Modifications} \label{sec.groext} 

To study other theories of gravity we can consider a change to the 
effective Newton's constant $G$ entering the above equations (remember 
the $G$ in the equations really means $G/G_{\rm Newton}$), or more 
generally some non-zero right hand side.  First we examine this as a 
generic change, and later treat a specific example within braneworld 
gravity. 

The deviation in $G$ from its Newtonian value caused by some time variation 
in $G$ can be viewed as a subset of a non-zero right hand side, since we 
may write a left hand side source term $XG(a)$ as the usual 
$XG_{\rm Newton}$ and add a term $X[G_{\rm Newton}-G(a)]$ to the right 
hand side.   Indeed, any difference between two cosmological models 
that only changes the source term, and keeps it linear in $g$, can be 
viewed as transforming the solution of the homogeneous differential 
equation for model 1 into a solution for model 2 of the effective 
inhomogeneous differential equation. 

Using a Green function method one can obtain a formal solution 
\beqa 
g(a_i,a)&=&\bar g(a_i,a)+\int_{a_i}^a du[\bar Q(u)-Q(u)]\,g(a_i,u) 
\nonumber\\ 
&\times&u^5H(u)\bar g(u)\bar g(a) 
\int_u^a dv\,\bar g^{-2}(v)\,v^{-5}H^{-1}(v)\,, \label{eq.green} 
\eeqa 
where $Q$ is the source term\footnote{Technically, $Q$ is the source term 
divided by the growth variable $g$, and also multiplied by $a^{-2}$ since 
Eq.~(\ref{eq.green}) uses a dependent variable of $a$ rather than $\ln a$. 
For example $Q=[(3/2)-(1/2)(\ln\om(a))'-(3/2)G\om(a)]a^{-2}$ 
in Eq.~(\ref{eq.groom}).}. 
The barred quantities represent some model 1 for which the solution 
is known (e.g.\ general relativity), and the integral 
gives the particular solution in the second model, 
for growth between any two scale factors (we can set $a_i=0$ to get the 
total growth up to some $a$). 

The solution can also be written as a recursion relation 
\beqa 
g(a_i,a)&=&\bar g(a_i,a)+\int_{a_i}^a du\,\bar g(a_i,u)\sum_{i=1}^\infty 
K_i(u,a) \label{eq.recurs} \\ 
K_{i+1}(u,a)&=&\int_u^a dx\,K_1(u,x)\,K_i(x,a) \\ 
K_1(u,a)&=&[\bar Q(u)-Q(u)]\,u^5H(u)\,\bar g(u)\bar g(a) \nonumber\\ 
&\times&\int_u^a dv\, 
\bar g^{-2}(v)\,v^{-5}H^{-1}(v). 
\eeqa 
This is particularly useful when considering small perturbations between 
models, e.g.\ when the gravitational coupling is slowly changing, 
as in the case of 
some scalar-tensor theories.  (Retaining only the first order term, $K_1$, 
is basically a Born approximation.) 

Another virtue of the Green function solution is the ability to see 
broad physical trends as the models change.  This follows the approach of 
\cite{linra,weissschneider} who considered the relation between distances 
as the cosmological model changed, including an analogous change in the 
theoretical framework (there in terms of allowing a clumpy universe).  
Here we consider the relation between growth factors.  If $\bar Q>Q$ then 
$K_i>0$ and so $g(a_i,a)>\bar g(a_i,a)$.  Thus we have a criterion 
for when the growth will be stronger, or when it will be more suppressed. 
With the expansion history fixed, 
the criterion $\bar Q>Q$ simply becomes $\bar G<G$; i.e.\ if the 
effective gravitational coupling is stronger than Newton's constant then 
the growth is enhanced.  For more elaborate modifications of gravity, 
a non-zero right hand side to the growth equation can contribute to 
$Q$ as well, but the prescription above still applies.

\section{A New Fitting Formula for Growth} \label{sec.fit} 

The general growth solutions of the previous section are formal, 
and while we saw that they can present generic physical insights they 
are somewhat cumbersome for application to cosmological models.  One 
might draw an analogy to trying to map the expansion history.  While 
one can calculate the expansion history in a specific model, say from 
a high energy physics scalar field potential, this is inefficient for 
comparison of the observations to many possible models.  Instead a 
useful approach is a model independent one, using a parametrization 
of the expansion history, for example in terms of the equation of state 
$w(z)$ value and variation: $w_0=w(z=0)$ and $w_a=(-dw/da)|_{z=1}$ 
(this is also similar to the inflationary power spectrum index and tilt 
parameters).  In this section we derive an analogous model independent 
parametrization of the growth history, putting it on equal footing 
with the cosmic expansion history. 

Rather than attempting to fit observations of growth history with 
an effective equation of state $w_{\rm grow}(z)$, it is better to 
render the physics appropriately: the expansion effects on the growth 
are described in terms of the standard expansion $w(z)$, and the 
gravitational modifications giving deviations from the expected 
growth history are treated as additional inputs.  Again, in the 
standard framework the expansion history completely determines the 
growth history.  Thus, we would like 
to write the growth history $g(a)$ as a function of an expansion history 
quantity plus a new, framework testing characteristic. 

Since the growth concerns matter density perturbations we take the 
expansion history in terms of the matter density history $\om(a)$. 
Two models with the same $\om(a)=\om a^{-3}/[H/H_0]^2$ for all redshifts 
will have the same expansion history.  So we look for a functional 
expression $g(\om(a))$.  One that works superbly well, as both a highly 
accurate approximation to the exact solution and as a simple characterization 
stimulating physical intuition, is 
\beq 
g(a)=e^{\int_0^a d\ln a\,[\om(a)^\gamma-1]}. \label{eq.gapx} 
\eeq 
Here $\gamma$ is our new parameter for the growth history, called the 
growth index, encompassing deviations in the theoretical framework. 
Models with identical expansion histories but different gravitational 
theories will possess different $\gamma$ parameters. 

\subsection{Accuracy Tests} \label{sec.gapxacc} 

First we establish the accuracy of the fitting formula, Eq.~(\ref{eq.gapx}), 
over a wide range of dark energy cosmologies.  Note that 
\cite{wangstein} (also see \cite{dweinberg,amendola}) 
has found that a similar formula provides estimations of the normalized 
growth factor at the accuracy level of about 1\%. 
However that approach normalized 
to the growth factor today (so it could not predict its value, and 
\cite{linjen} showed that $g(z)/g(0)$ varies by only a few percent 
innately between models) and fixed 
the growth index.  Furthering the pioneering work of 
\cite{wangstein,dweinberg,amendola} we can remove both those restrictions and 
obtain an order of magnitude better accuracy. 

In terms of the expansion history dark energy equation of state, 
within general relativity, we find excellent fits, to better than 0.2\%, using 
\beq 
\gamma=0.55+0.05[1+w(z=1)]. \label{eq.gamma} 
\eeq 
Employing the value of the equation of state evaluated at $z=1$ allows 
simple treatment of dynamical models where the equation of state varies 
with redshift, as it generically does. 

For the cosmological constant case, the fitting formulas Eqs.~(\ref{eq.gapx}) 
and (\ref{eq.gamma}) reproduce the exact growth history for any redshift 
(including the 
total growth to the present) to better than 0.05\% over the range 
$\om\in[0.22,1]$!  It remains accurate to better than 1\% all the way 
down to $\om=0.01$.  

Models with equation of state $w=-0.8$ ($-0.5$) have accurately fit 
growth histories to within 0.2\% (0.4\%) for $\om\in[0.2,1]$.  A dynamical 
model such as SUGRA with $w_0=-0.82$, $w_a=0.58$ is fit to within 0.25\%. 
Models with equations of state $w<-1$ are similarly well approximated. 
When $w=-1.2$ ($-1.5$), the fit is good to 0.3\% (0.5\%).  If we are 
willing to slightly modify the simplest fit of Eq.~(\ref{eq.gamma}) to 
\beq 
\gamma=0.55+0.02[1+w(z=1)]\quad {\rm for\ } w<-1, \label{eq.gammam1}
\eeq 
for the phantom models $w<-1$, then we achieve an astonishing 0.05\% 
accuracy for these fits.  
(Note that the fitting function of \cite{heath,peebles80}, also 
containing a single 
integral, is accurate to only 5\% for models with $w=-0.8$ or $-1.2$ and 
$\om=0.3$.) 

While impressive in accuracy, the growth function fitting formula's 
primary purpose is not a fit as such (the exact solution requires only 
solving a second order differential equation), but rather its usefulness in 
physical intuition and in parametrizing modifications of the Einstein growth 
equation beyond the expansion behavior (just as $w(z)$ parametrizes 
modifications of the Friedmann expansion equation).  The fitting function 
provides us access to the acceleration physics that exists beyond what 
the expansion history sees. 

\subsection{Example: Braneworld Gravity} \label{sec.bw} 

The growth of matter perturbations in gravitational theories beyond 
general relativity is not well developed.  Here we consider one theory 
that has been shown to be self-consistent 
\cite{dgp,ddg,lue,gruzinov,gabadadze}, the DGP \cite{dgp,ddg} 
braneworld theory of gravity.  In this theory gravity has infrared 
modifications due to spacetime possessing a large extra dimension 
(making our view a 4D brane within a 5D bulk), causing a weakening of 
gravity on large scales approaching the Hubble scale. 

The expansion history for this braneworld theory follows from the 
modified Friedmann equation, 
\beq 
H^2-H/r_c=(8\pi/3)\rho, 
\eeq 
where $r_c=H_0^{-1}/(1-\om)$ is the crossover distance, 
related to the 5D Planck mass. 
Equivalently the expansion history has an effective equation of state 
\beq 
%w(a)=-1+\sqrt{\om(a)/[4+a^3(1-\om)^2/\om]}. 
w(a)=-[1+\om(a)]^{-1}, 
\eeq 
as noted elegantly by \cite{lue}. 
The braneworld expansion history can be well 
approximated by 
a simple scalar field model with $w_0=-0.78$, $w_a=0.32$.  Indeed these 
two very different physical origins for the acceleration agree in 
distance measurements to within 0.5\% (0.01 mag) out to $z=2$. 

\begin{figure}
\begin{center}
\psfig{file=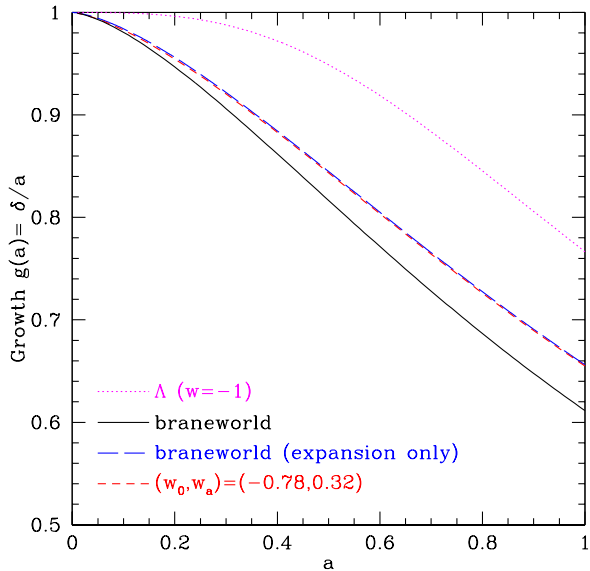,width=3.2in}
\caption{The growth history is shown for 
an extra dimensional braneworld model (long dashed, blue curve) 
and a quintessence model with $w_0=-0.78$, $w_a=0.32$ (short dashed, red), 
having nearly identical expansion histories.  When proper account is taken 
of the effects 
of altered gravity on the braneworld growth history (solid, black curve) 
this allows distinction of these models. 
The expansion history can in turn rule out a quintessence model 
degenerate with the solid curve. 
}
\label{fig.grobw} 
\end{center}
\end{figure}

Information from the growth history, however, as stated before can 
break this degeneracy in the nature of the acceleration physics. 
Comparing the braneworld model with a scalar field model with an 
identical expansion history shows deviations in the present growth 
factor of 7\%.  Figure~\ref{fig.grobw} illustrates how the growth 
history depends on both expansion history and the gravitational 
framework.  Taking into account only the expansion history in the 
growth equation, the braneworld and scalar field models appear to 
have the same growth history.  However, proper treatment of the 
gravitational modifications inherent in the braneworld scenario 
separate these models.  This has been pointed out as well in 
\cite{lue,song,knoxsongtyson}.  

In the linear power spectrum the deviation 
ranges from 4\% at $z=2$ to 15\% today.  While a scalar field model 
that matched the modified growth of the braneworld model is possible, 
it in turn can be distinguished through the expansion history.  We 
see that expansion measurements and growth measurements work in 
important complementarity to reveal the nature of the new physics. 

\begin{figure}
\begin{center}
\psfig{file=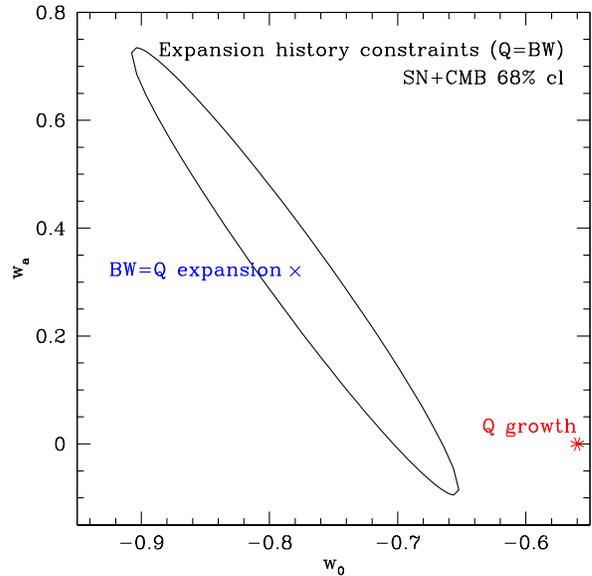,width=3.2in}
\caption{Expansion history and growth history constraints on the 
dark energy equation of state parameters can test the theoretical 
framework by looking for inconsistent results.  The blue cross gives 
the best fit for the expansion history of a quintessence (Q) universe 
matching the braneworld (BW) scenario, but the red star gives 
the best fit for the growth history to a quintessence model, assuming 
general relativity.  The black ellipse shows the constraints at 68\% 
confidence level from next generation data composed of SNAP supernovae 
data and Planck CMB last scattering distance measurement.
}
\label{fig.gapw0wa} 
\end{center}
\end{figure}

Figure~\ref{fig.gapw0wa} demonstrates this synergy explicitly.  Supposing 
the universe was described by a braneworld model with $\om=0.28$, distance 
measurements of the quality of the proposed Supernova/Acceleration Probe 
(SNAP; \cite{snap}) supernova data set, together with a 0.7\% measurement 
of the 
distance to the CMB last scattering surface from Planck \cite{planck}, 
would provide the constraints (at 68\% confidence level, marginalizing over 
other parameters such as $\om$) in the dark energy 
equation of state parameter space shown by the ellipse.  The best fit 
for growth measurements would concur with the solution using braneworld 
gravity equations, showing the consistency of the data with this 
model.  However, if the growth equation employed general relativity, 
the best fit would lie at the red star, $w_0=-0.56$, $w_a=0$, clearly 
inconsistent.  The comparison of expansion and growth histories reveal 
a breakdown of the theoretical framework, this discrepancy alerting us 
to a possible modification of gravity (or experimental systematic errors). 

Of course if the measurements were too coarse and imprecise, we would not 
necessarily have noticed a statistically significant discrepancy.  The 
braneworld model expansion is compatible with an expansion history of 
a constant $w=-0.71$ model, to 1\% in distance out to $z=1.7$.  So the 
expansion history measurements find a ``distance'' in equation of state 
space of $\Delta w_0=0.15$ between the effective scalar field model from 
the expansion history and that from the growth history.  Conversely, 
the expansion history of $(w_0,w_a)=(-0.56,0)$ can be fit by $(-0.63,0.32)$, 
so the ``distance'' from the expansion fit to the braneworld model of 
$(-0.78,0.32)$ is again $\Delta w_0=0.15$.  This suggests 
that for a $3\sigma$ detection of framework inconsistency we should 
strive for experiments that provide an uncertainty of $\sigma(w_0)<0.05$. 

Likewise one can estimate from the different orientations of the 
expansion history and growth history constraints in the $w_0-w_a$ plane 
that the precision of measurements on $w_a$ should be $\sigma(w_a)<0.2$. 
This comparison of growth to expansion provides one of the only ways of 
putting an absolute scale on the measurement precision that should be 
striven for in experiments to reveal the nature of the accelerating 
physics -- a significant breakthrough (see also \cite{caldlin}).  
This is somewhat dampened by 
the realization that this scale is particular to the braneworld 
scenario.  Note that the relative precisions between $w_0$ and $w_a$ 
obey the relation 
\beq 
\sigma(w')\equiv\sigma(w_a)/2\approx 2\sigma(w_0), 
\eeq 
found in \cite{caldlin,caldlin2}, though that analysis 
was within the scalar field context.  This relation signifies that a 
precise measurement of $w_0$ is of limited use without concomitant 
constraint on $w_a$, since a sufficiently different $w_a$ can spoof 
$w_0$.  That is, the uncertainty in seeing a discrepancy will be 
dominated by the largest error among the two equation of state parameters. 

To move beyond a mere alarm that there is an inconsistency, we need 
to employ the growth parametrization of Eq.~(\ref{eq.gapx}) to obtain a 
quantitative measure of the deviation from the growth behavior predicted 
by the expansion history measurements.  We find that the fitting formula 
works for the braneworld scenario including gravitational modifications, 
using a growth index $\gamma=0.68$ (note that the pioneering paper of 
\cite{lue} indicated the equivalent of $\gamma=2/3$).  In 
fact, the growth history using the fitting function Eq.~(\ref{eq.gapx}) 
and $\gamma=0.68$ matches the exact solution to within 0.2\% (for 
$\om\ge0.2)$.  

The approximation 
of a single growth parameter beyond the expansion history effects on the 
growth can be validated by asking what values of $\gamma$ as a function of 
redshift reproduce the exact solution.  For the case $\om=0.28$, the 
(now) function $\gamma(z)$ ranges between 0.665 at $z=0$ to 0.683 at 
$z=1$ to 0.687 at high redshift.  (The constancy of $\gamma$ with 
redshift holds even 
better for quintessence models.)  This, as well as the excellent fit to 
the growth function, justifies the use of a single parameter $\gamma$, 
the growth index. 

\begin{figure} 
\begin{center} 
\psfig{file=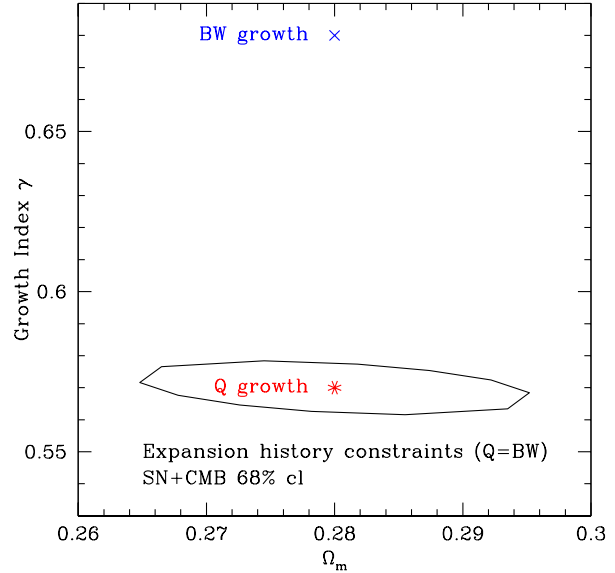,width=3.2in} 
\caption{While Fig.~\ref{fig.gapw0wa} showed that expansion history 
and growth history constraints on the 
dark energy equation of state parameters could test the theoretical 
framework by looking for inconsistent results, here we see quantitative 
measures of framework breaking by gravitational modification of the 
growth index $\gamma$.  The red star gives the best fit for a quintessence 
(Q) model matching the expansion history of the 
braneworld (BW) scenario, but the blue cross gives the true result for the 
braneworld growth history.  The black ellipse shows the constraints at 68\% 
confidence level from next generation data composed of SNAP supernovae 
data and Planck CMB last scattering distance measurement, marginalized 
over the equation of state parameters $w_0$, $w_a$.
}
\label{fig.gapxom} 
\end{center} 
\end{figure}

With this model independent parametrization in hand, we can obtain 
quantitative measures of the deviations between models, even those that 
involve gravitational modifications.  To the cosmological parameters 
$\om$, $w_0$, and $w_a$, we add the growth index $\gamma$ and can plot 
the resulting parameter estimation uncertainties, marginalizing over 
subsets of parameters.  Figure~\ref{fig.gapxom} illustrates 
an example. 

The growth index $\gamma$ that would fit the braneworld growth history 
is clearly distinct from the values allowed by a scalar field model that 
matches the expansion history.  The ``distance'' in $\gamma$ is 0.11; 
to attain the value $\gamma=0.68$ would require, by extrapolation of 
Eq.~(\ref{eq.gamma}), $w=+1.6$!  A $3\sigma$ distinction of the 
framework breaking would need a measurement with precision 
$\sigma(\gamma)=0.037$ (marginalized over the other parameters).  This 
corresponds roughly to a 2\% measurement of the growth history.  Indeed, 
this is in good agreement with the results of Fig.~\ref{fig.grobw}, 
which showed growth deviations between the two models with identical expansion 
histories at up to the 7\% level, so the same $3\sigma$ criterion leads 
to $\sim2$\% precision. 

\subsection{Coupling of Matter and Dark Energy} \label{sec.couple} 

While the primary purpose of the formalism here is to test the 
gravitational framework, in the case of a physical dark energy there 
can enter microphysical effects.  These can include spatial perturbations 
to the dark energy or coupling to the matter component.  We leave the 
first of these to future work, but note that growth probes involving 
correlations of large scale structure with the CMB might play a role 
(e.g.\ \cite{coraisw,bacciisw,gold}).  Here we consider whether the fitting 
formula and growth index approach remain valid in the presence of coupling. 
Without a microphysical theory these are necessarily toy models, and we 
only consider the effects on matter growth, neglecting early universe 
or fifth force constraints. 

Interaction between matter and a dark energy component is treated through 
a coupling of the evolution equations: 
\beq 
\dot\rho_i=-3H(1+w_i)\rho_i+\Gamma_i(a,\rho_m,\rho_{de}). \label{eq.rhodot}
\eeq 
We have considered the cases $\Gamma_m=-\Gamma_{de}=\alpha\rho_m$, 
$\alpha\rho_{de}$, and 
$\alpha a^n H$, the slinky inflation model \cite{slinky}, and the undulant 
universe model \cite{undulant}.  Note the undulant universe is a special 
case of the slinky model, without coupling, and is ruled out by having a 
very low growth factor ($g_0=0.03$).  

All these models follow the growth index 
formalism if the coupling is not too strong.  As the coupling increases 
(e.g.\ if the dimensionless coupling $\alpha'=\alpha/\rho_m(0)\ga0.5$ in the 
$\Gamma_m=\alpha H$ case), this will start to break down because the 
equation of state of matter begins to deviate significantly from zero 
(plus in some cases the high redshift universe is not matter dominated). 
As a simple example, consider the decaying matter scenario where 
$\Gamma_m=-\alpha\rho_m$.  This was treated in detail in \cite{turner85}, 
and the matter equation of state is $\alpha/(3H)$ \cite{lin88cos}. 
Generically, a coupling of the form $\Gamma_i=\alpha\rho_i$ will change 
the equation of state of component $i$, defined by the effective conservation 
equation $\dot\rho_i=-3H(1+w_{\rm eff})\rho_i$, from $w_i$ to 
$w_{\rm eff}=w_i-\alpha/(3H)$ 
(providing a way for a $w>-1$ component to look phantom, $w<-1$). 

Further research into the effects of coupling on growth of perturbations 
is underway \cite{luke} (also see \cite{amenskew}).  Lack of a 
consistent microphysical theory is 
the major obstacle.  For example, is the ``new'' energy density in a 
component distributed uniformly or in the same spatial distribution as 
the component it came from?  Issues of evolution from that point by 
clumping or free streaming make rigorous calculation complicated.

\section{Conclusions} \label{sec.concl} 

To reveal the physical origin of the acceleration of the universe, 
both probes of the expansion history (such as the distance-redshift 
relation) and of the growth history (such as weak gravitational lensing 
measurements involving the mass power spectrum) are required.  While 
the two types of probes in synergy give enhanced constraints on the 
effective dark energy equation of state, in comparison they can test 
the theoretical framework of cosmology and general relativity. 

The growth history of mass in the universe follows the source and friction 
term behaviors governed by the expansion.  Deviations from this reveal 
a breakdown of the framework such as from modification of gravity.  By 
rendering the growth function in a physically appropriate manner, 
separating the expansion effects from framework extensions, we 
presented here a new, physically intuitive and highly accurate 
(0.05-0.2\%) fitting function, Eq.~(\ref{eq.gapx}), for the 
linear growth of perturbations in generalized cosmologies. 
This allows model independent quantification of gravitational modifications 
in terms of a new parameter, the growth index $\gamma$. 

This research suggests a new paradigm for understanding the nature of 
the acceleration physics: accurate measurement of expansion and growth 
separately, for example through Type Ia supernovae and weak gravitational 
lensing.  A useful, model independent, quantitative parameter set was 
shown to be the equation of state value $w_0$ and variation $w_a$ and 
the growth index $\gamma$.  In the specific worked case of comparing 
an extra dimensional braneworld scenario with scalar field physics in 
general relativity, the desired measurement precisions should be of 
order $\sigma(w_0)\le0.05$, $\sigma(w_a)\le0.2$, $\sigma(\gamma)\le0.04$. 
These should be technically feasible and should be within the reach of 
next generation experiments such as the Joint Dark Energy Mission. 

The formalism presented here has further applications for future 
investigation, such as seeing the effect of perturbations in a physical dark 
energy component, couplings between dark energy and matter, and 
scalar-tensor gravity.  To reveal the nature of the new physics responsible 
for the universe-shaking acceleration, we will require a comprehensive 
suite of cosmological probes.  The significance of the discoveries is 
so great that every robust method is needed to strengthen the accuracy, 
and the confidence in our understanding.  With clear measurements of 
the cosmic expansion history and the cosmic growth history together 
we can learn if nothing weighs something, if nowhere is somewhere, 
or even more unexpected insights.

%\section*{Acknowledgments} 
\begin{acknowledgments} 

This work has been supported in part by the Director, Office of Science,
Department of Energy under grant DE-AC02-05CH11231.  I thank Dragan 
Huterer, Arthur Lue, Roman Scoccimarro, David Weinberg, and Martin 
White for useful conversations, and the referee for careful reading and 
useful suggestions.  

\end{acknowledgments}

\appendix*

\section{Inhomogeneities and Cosmic Expansion} \label{sec.apxjlw} 

We have taken the cosmic growth history to not ``back react'' on 
the cosmic expansion history.  That is, the global homogeneous expansion 
independent of the growth of matter structure is a valid treatment. 
This is a topic of great interest and 
comment; here we simply present a brief summary of the dependence of 
the metric on gravitational potentials and the lack of large contributions 
by gravitational potentials (in particular no infrared divergence) to 
the cosmic expansion. 

The approach taken by \cite{jlw1,jlw2} is a straightforward calculation 
to obtain the metric of a realistically inhomogeneous universe.  In 
particular, it did not rely on any averaging procedure, rather a 
harmonic decomposition of the perturbations.  The second key aspect 
was no a priori assumption on the size of matter density fluctuations; 
rather it used a post-Newtonian parametrization, essentially a weak 
field, slow motion expansion.  This followed work of Futamase 
\cite{futamase1,futamase2,futamase3} and can be traced back to the 
mean field theory, 
or two length scale, approach of Chandrasekhar \cite{chandra}. 

For potentials parametrized by an amplitude $\epsilon^2\ll1$, and 
characteristic length scale $\kappa$, the slow motion or, more physically, 
the shear condition $\epsilon^2/\kappa\ll1$ applies.  Violation of this 
condition leads to ray crossing in light propagation (see \cite{jlw2,linsky}) 
and eventually relativistically moving matter structures, contrary to 
observations of our universe.  Landau \& Lifshitz \cite{landl} 
pointed out that the dominant first order effect on the cosmic expansion 
entered at what they called pseudotensor order: $\epsilon^4/\kappa^2$. 
Thus the shear condition ensures that the expansion is insignificantly 
affected, and conversely a significant back-reaction of inhomogeneities on the 
expansion would generically lead to visible anisotropies. 

However, here we concentrate on the post-Newtonian gravitational 
potentials, and modification of the Newton-Poisson equation relating  
the potentials to the matter density distribution.  The general solution 
obtained by \cite{jlw1,jlw2} was 
\beqa 
\phi(\eta,\vec x)&=&-\frac{4\pi}{3}\int_{\eta_0}^\eta \frac{du}{a'(u)} 
\nonumber \\ 
&\times& \int d^3\vec y\,a^3(u)\,\delta\rho(u,\vec y)\,
{\cal G}(u,\eta, \vec x,\vec y), 
\eeqa 
plus an initial condition term.  The Green function is 
\beqa 
{\cal G}(u,\eta,\vec x,\vec y)&=&[a(u)/a(\eta)]\,[4\pi C(u,\eta)]^{-3/2} 
\nonumber \\ 
&\quad& \times\ e^{-|\vec y-\vec x|^2/[4C(u,\eta)]} \\ 
C(u,\eta)&=&(1/3)\int_u^\eta dv\,(a/a'), 
\eeqa 
where a prime denotes a derivative with respect to the conformal time $\eta$. 

These expressions show that there is no divergence of the potential 
or its derivatives in the presence of inhomogeneities.  In contrast, 
the post-Newtonian Green function solution, while matching the Newton-Poisson 
equation on small scales, shows an exponential suppression of 
the potential as one approaches horizon scales.  These limits are 
treated in detail in \cite{jlw2}, and the physical problem is shown to 
be closely analogous to the displacement probability distribution for 
isotropic random walks, and for diffusion in a uniform medium.

\end{document}